\documentclass[]{aastex7}

\newcommand\fnu{\(F_\nu\)}
\newcommand\snu{\(\sigma_\nu\)}
\newcommand\tnuc{\(t_{\rm nuc}\)}
\usepackage{amsmath}

\begin{document}

\title{Physical Conditions for Synthesis of Sc, Ti, and V in Neutrino-driven Supernovae }

\author[orcid=0009-0005-8843-8262,sname='Hatami']{Ryota Hatami}
\affiliation{Astronomical Science Program, Graduate Institute for Advanced Studies, SOKENDAI 2-21-1 Osawa, Mitaka, Tokyo 181-8588, Japan}
\affiliation{National Astronomical Observatory of Japan, National Institutes of Natural Sciences, 2-21-1 Osawa, Mitaka, Tokyo 181-8588, Japan}
\email[show]{ryota.hatami@grad.nao.ac.jp}  

\author[orcid=0000-0001-8537-3153,sname='Tominaga']{Nozomu Tominaga}
\affiliation{National Astronomical Observatory of Japan, National Institutes of Natural Sciences, 2-21-1 Osawa, Mitaka, Tokyo 181-8588, Japan}
\affiliation{Astronomical Science Program, Graduate Institute for Advanced Studies, SOKENDAI 2-21-1 Osawa, Mitaka, Tokyo 181-8588, Japan}
\affiliation{Department of Physics, Faculty of Science and Engineering, Konan University, 8-9-1 Okamoto, Kobe, Hyogo 658-8501, Japan}
\email{nozomu.tominaga@nao.ac.jp}

\author[orcid=0000-0002-8967-7063,sname='Yoshida']{Takashi Yoshida}
\affiliation{Yukawa Institute for Theoretical Physics, Kyoto University, Kyoto 606-8502, Japan}
\email{fakeemail3@google.com}

\author{Hideyuki Umeda}
\affiliation{Department of Astronomy, Graduate School of Science, The University of Tokyo, Tokyo 113-0033,
Japan}
\email{fakeemail4@google.com}

\author[orcid=0000-0003-0304-9283,sname='Takiwaki']{Tomoya Takiwaki}
\affiliation{Astronomical Science Program, Graduate Institute for Advanced Studies, SOKENDAI 2-21-1 Osawa, Mitaka, Tokyo 181-8588, Japan}
\affiliation{National Astronomical Observatory of Japan, National Institutes of Natural Sciences, 2-21-1 Osawa, Mitaka, Tokyo 181-8588, Japan}
\email{fakeemail5@google.com}




\begin{abstract}

We present the results of simulations of nucleosynthesis in a core-collapse supernova (CCSN) including the neutrino process. Using the Si layer of \(13M_\odot\) zero-metal progenitor as the initial composition, we calculate the nucleosynthesis by adopting the temperature, density, neutrino flux, and duration of nucleosynthesis as arbitrary parameters and compare the results with the observed abundances ratio of Sc, Ti, and V in very metal-poor (VMP) stars taken from the Stellar Abundances for Galactic Archaeology (SAGA) database. As a result, for the first time, we identify the quantitative requirements on local physical conditions. To reproduce the abundances ratios in the VMP stars, the explosive nucleosynthesis should take place under the neutrino exposure, which is time integration of neutrino flux, of \(\sigma_\nu\sim 10^{35}\,\mathrm{erg~cm^{-2}}\) and temperature of \(2.0\,\mathrm{GK}\leq T \leq 3.2\,\mathrm{GK}\). The dependence on the density and each value of the neutrino flux and the duration of nucleosynthesis is weak. We also discuss whether the quantitative requirements are realized during the explosion. Although the requirements are difficult to be realized in the one-dimensional simulations, the non-monotonic thermal evolution shown in recent three-dimensional simulations may satisfy them. Because the evolution is likely caused by turbulent motion stemming from the initial asphericity of the progenitor, it is important to calculate the long-term three-dimensional supernova explosion of multi-dimensional metal-free progenitor models and follow the nucleosynthesis self-consistently. 



\end{abstract}

\keywords{\uat{Nucleosynthesis}{1131}; \uat{Explosive nucleosynthesis}{503}; \uat{Core-collapse supernovae}{304}; \uat{Supernova dynamics}{1664}; \uat{Stellar abundances}{1577}}


\section{Introduction} 
Core-collapse supernovae (CCSNe) are the explosions of massive stars triggered by the gravitational collapse. During the evolution of a massive star from its formation to gravitational collapse, various heavy elements up to Fe are synthesized in the stellar interior. Explosive nucleosynthesis takes place at the explosion because of the high-temperature and high-density environment behind the shock wave \citep{Hoyle54}. Metals heavier than carbon are ejected into the universe by the CCSNe. This drives the chemical evolution of the universe. Therefore, the understanding of the evolution of massive stars and CCSNe is an important issue in astronomy.  

One of the plausible explosion mechanisms for a CCSN is the neutrino-heating mechanism \citep{BW85}. In this mechanism, a shock wave is produced at the core bounce, but stalls because of the ram pressure of the infalling materials and the stalled shock wave is heated by the absorption of neutrinos, i.e., \(n(\nu_e,\,e^-)p\) and \(p(\bar{\nu}_e,\,e^+)n\), emitted from a proto-neutron star. Although this mechanism is promising, it has been shown that a CCSN cannot explode in an ab initio calculation solving the full neutrino transports under spherical symmetry \citep{Sumiyoshi05}. 

Recent numerical simulations have demonstrated that multi-dimensional effects, such as a convection \citep[e.g.,][]{Burrows95,Janka96} and standing accretion shock instability \citep[SASI;][]{Blondin03, Foglizzo06, Ohnishi06, Foglizzo07, Iwakami08, Iwakami09, Couch13eos}, are crucial for successful explosions \citep[e.g.,][]{Takiwaki12, Suwa13, Nakamura15, Mueller15, Bollig21, Burrows24, Nakamura25}.
These multi-dimensional effects enhance the advection time resulting in more efficient heating and aid in the revival of the shock wave. The revived shock wave penetrates the stellar mantle, emerges from the stellar surface, and expands into the universe. The expanding ejecta is observed as a CCSN. 

The most important properties of CCSNe are the explosion energy and the ejected amount of \(\mathrm{^{56}Ni}\). Recent transient surveys have revealed that the typical explosion energy and \(\mathrm{^{56}Ni}\) mass of type II supernovae range from \(\sim0.6 \times 10^{51}~{\rm erg} = 0.6~\mathrm{B}\) (\(1~\mathrm{B}:=10^{51}~\mathrm{erg}\)) to a few Bethe (\(\gtrsim 1~\mathrm{B}\)) and \(\sim10^{-3}\text{--} 10^{-1}M_\odot\), respectively \citep{Anderson19,Martinez22a, Martinez22b, Martinez22c,Subrayan23,Silva24}. 
Many multidimensional simulations have tended to yield explosion energies or 
\(\mathrm{^{56}Ni}\) masses lower than observed \citep[e.g.,][]{Bruenn13,Nakamura15,Nakamura25,Summa16,Burrows19}.
These simulations were terminated at relatively early post-bounce times, likely before the explosion energy had converged.
Recently, long-term simulations have shown that several models have reached the explosion energy of \(1\,\mathrm{B}\) and the \(\mathrm{^{56}Ni}\) mass of \(0.1M_\odot\) \citep{Bollig21, Sieverding23a, Burrows24,Wang24a}.
However, the predicted observables are sensitive to the numerical methodology and microphysics \citep[e.g.,][]{Just18,Nagakura19}. Furthermore, the outcome may change depending on the treatment of neutrino flavor conversion \citep{Ehring23,Mori25}. Thus, despite recent progress, the available models are still too limited to draw definitive conclusions, and our understanding of the explosion mechanism remains under development.
Moreover, the explosion energy derived from a numerical simulation is the consequence of all relevant processes in a CCSN. Thus, it is difficult to extract the information that allow us to scrutinize each process involved in the explosion mechanism of a CCSN.



Information on each physical process is contained in the explosive nucleosynthesis yield of a CCSN because explosive nucleosynthesis takes place during the explosion and primarily depends on the local thermal history and neutrino exposures of materials. 
Two classes of astronomical objects are frequently used to observe the nucleosynthesis of CCSNe. The first classes are supernova remnants (SNR), which is a nebula-like object left behind for several tens to ten thousand years after a supernova explosion. The other classese are very metal-poor (VMP) stars with low metallicity of [Fe/H]\footnote{Here, \([\mathrm{X/Y}] = \log\left(N_\mathrm{X}/N_\mathrm{Y}\right)-\log\left(N_\mathrm{X}/N_\mathrm{Y}\right)_\odot\), where \(N_\mathrm{X}\) and \(N_\mathrm{Y}\) are the abundances of elements X and Y respectively.}~\(\leq-2\), which is formed from a mixture of pristine gas and metals ejected by CCSNe in the early universe.

Recent X-ray observation of Cassiopeia A (Cas A) detected the emission lines of Ti, Cr, and Mn. The abundance ratios among them demonstrated the necessity of the neutrino process in the explosion of Cas A \citep{Sato23}. 
The amount of radioactive isotope \(\mathrm{{^{44}Ti}}\) and also \(\mathrm{^{56}Ni}\) are compared with nucleosynthesis calculations based on long-term three-dimensional simulations \citep{Sieverding23a, Wang24a}. 
Although these are observational evidences that the CCSN ejecta interacts with neutrinos during the explosion, it is difficult to quantitatively constrain the physical quantities related to the nucleosynthesis and neutrino process because SNR observations have difficulty detecting the lines of other elements with low abundance.


In contrast to the observations of CCSNe and SNRs, spectroscopic observations of the VMP stars make it possible to measure elements with low abundance such as odd-Z elements \cite[e.g.][]{Roederer14}, although a CCSN explosion model with a metal-free progenitor is needed for the comparison. The metal abundances of VMP stars and their evolution are mostly explained by the nucleosynthesis yields of CCSNe \citep[e.g.][]{nkt13} and the galactic chemical evolution simulations \citep{Kobayashi20}. However, there are longstanding problems that some elements cannot be reproduced by them. Solutions to these discrepancies will give us a clue to understanding the interaction between neutrinos and matters during the explosion and thus the explosion mechanism of CCSNe. 



We focus on Sc, Ti, and V, which are underproduced in the nominal nucleosynthesis yields of CCSNe \citep[e.g.][]{nkt13}. Recent observations have revealed positive correlations among the abundance ratios [(Sc, Ti, and V)/Fe] \citep{Sneden_2016}. These correlations indicate that Sc, Ti, and V are synthesized in the same layer, which is different from the Fe synthesis layer. 
Although there have been several suggestions on ways to enhance the ratios, such as high-entropy environments \citep[e.g.][]{Maeda03, Umeda05, Magkotsios10, Tominaga14, Leung23, Leung24} and neutrino processes that are caused by neutral current reactions such as \(\nu + (Z,\,A)\rightarrow \nu^\prime + (Z,\,A-1) + n\) and subsequent neutron or proton capture reactions, \citep[e.g.][]{Woosley90, Yoshida08}, 
theoretical models have difficulty to consistently reproduce the abundance ratios among Sc, Ti, and V or their scatters. For example, a high neutrino luminosity in the neutrino processes, which reproduces the Sc abundance, results in the overproduction of V \citep{Yoshida08}. 
Scatters of [Sc/Ti] and [V/Ti] are much larger than the observed ratios, and an integration with Salpeter initial mass function cannot reproduce [Sc, Ti/Mg] \citep{HW10}.\footnote{Some integrations of the yields over specific Gaussian initial mass functions can fit to these ratios in \citet{Cayrel04}.} 
Although \citet{Grimmett18} showed that the abundance ratios among these three elements can be reproduced in \(15M_\odot\) and \(20M_\odot\) progenitor with the explosion energy of \(<10\,\mathrm{B}\) with the same progenitor models as \citet{HW10}, their models cannot explain the scatters of [Sc/Ti] and [V/Ti] and their treatment can underestimate the fallback and overestimate the mass of these elements in ejecta. 
Another study of explosions of metal-free progenitors with fast rotations ($\sim150-700$~km~s$^{-1}$) can reproduce the abundance ratios among Sc, Ti, and V but underproduce Mg, resulting in the [X/Mg] higher than observations \citep{Roberti24}. 

In this paper, we investigate the local physical conditions needed to reproduce the abundances ratios among Sc, Ti, and V because the observed correlation indicates that Sc, Ti, and V are synthesized in the same layer. Although the abundance ratios between elements synthesized in the different layers (e.g., [Ti/Fe]) are also important for understanding supernova nucleosynthesis, they include other ambiguities such as mixing and hydrodynamical structure during the explosions and require long-term self-consistent multi-dimensional numerical simulations of CCSNe. Instead, we derive the local physical conditions that reproduce the abundance ratios among these three elements (i.e., [Sc/Ti] and [V/Ti]) consistently and discussed the feasibility of realizing the required conditions in the numerical simulations. 

This paper is structured as follows. We introduce the models, methods, and observational data in Section \ref{sec:models}. In Section \ref{sec:results}, we demonstrate the necessity of neutrinos and derive the physical conditions required to reproduce the abundance ratios among Sc, Ti, and V in the VMP stars. Then, we discuss the feasibility of realizing such conditions in CCSNe in Section \ref{sec:discussion}. Finally, we present our conclusions in Section \ref{sec:conclusion}.

\section{Models and Methods} \label{sec:models}

\subsection{Model}
\label{sec:mod}

\begin{figure}
    \centering
    \includegraphics[width=0.7\linewidth]{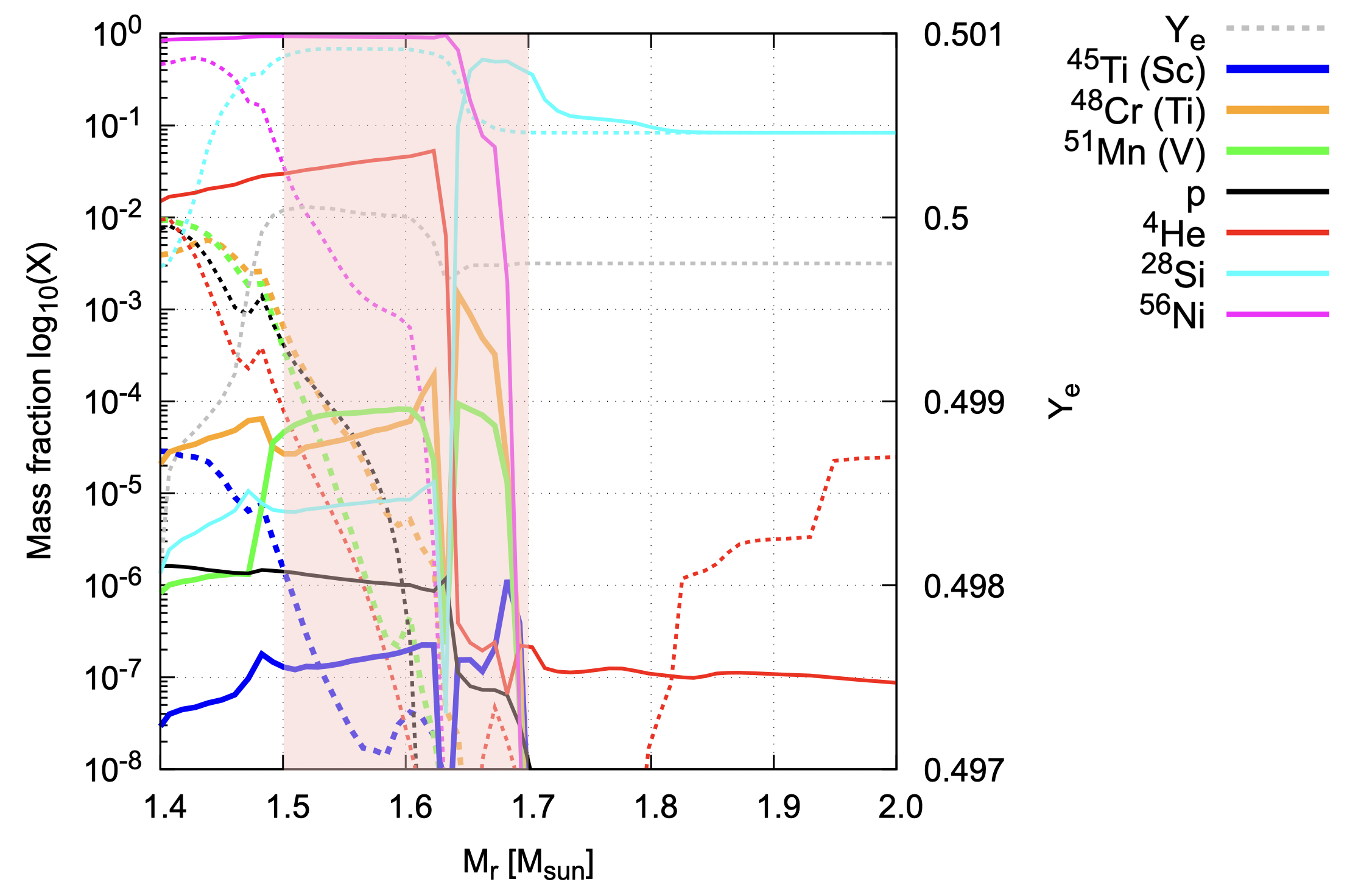}
    \caption{Abundance distribution of a zero-metal \(13M_\odot\) progenitor model (dashed line) and the ejecta after explosive nucleosynthesis (solid line; a 1D thermal bomb model with \(1~\mathrm{B}\), \citealt{Tominaga07}). The colors represent the species of nuclei. The gray dashed line indicates \(Y_e\) distribution of the progenitor. The shaded region indicates the initial composition. 
    }
    \label{fig:initial_composition}
\end{figure}

We adopt a progenitor model with the zero-age main-sequence mass of \(13M_\odot\) and the zero metallicity \(Z=0\) \citep[Figure \ref{fig:initial_composition},][]{Umeda00} to simulate the normal supernovae contributing to the abundance ratios of VMP stars \citep[e.g.,][]{Tominaga07}. 
One-dimensional explosive nucleosynthesis calculations with the same progenitor and explosion energy of \(10^{51}\)~erg, initiated by a thermal bomb, shows that Sc (\(^{45}\)Ti), Ti (\(^{48}\)Cr), and V (\(^{51}\)Mn) are mainly synthesized by Si burning at \(<1.7M_\odot\) \citep{Tominaga07}. Furthermore, the layer inner than \(1.5M_\odot\) is close to the iron core and the materials in the layer are likely to fall onto the central remnant. Therefore, we investigate the nucleosynthesis in the layer of \(1.5\,\mathchar`-\,1.7M_\odot\) of the progenitor model. 
The layer mostly consists of Si and O and has electron fraction (\(Y_e\)) of \(\sim0.5\).
We calculate the final abundance ratios by integrating the masses of elements over this mass range and adopt the solar abundance in \citet{Asplund09}.

To constrain the local physical conditions that reproduce the abundance ratios among Sc, Ti, and V, we do not assume any explosion model or do not calculate hydrodynamic simulations, but instead perform nucleosynthesis calculations by adopting the temperature \(T\), density \(\rho\), neutrino flux \(F_\nu\), and duration of nucleosynthesis \(t_{\rm nuc}\) as independent parameters. 
We adopt constant \(T\), \(\rho\), and \(F_\nu\) with 
certain ranges and steps, as listed in Table~\ref{table:parameters}.  
The temperature is varied in step of \(0.1\,\mathrm{GK}\) from \(1.0\,\mathrm{GK}\) to \(7.0\,\mathrm{GK}\) because explosive nucleosynthesis is occurred in this temperature range. The density is varied by one order of magnitude from \(10^5\,\mathrm{g~cm^{-3}}\) to \(10^8\,\mathrm{g~cm^{-3}}\), which corresponds to the typical density of the CCSN ejecta during the explosive nucleosynthesis. 
The neutrino flux changes by one order of magnitude from \(2.65\times10^{31}\,\mathrm{erg~cm^{-2}~s^{-1}}\) to \(2.65\times10^{37}\,\mathrm{erg~cm^{-2}~s^{-1}}\), which corresponds to the radii from \(10^7~\)cm to \(10^{10}\) cm, assuming a typical neutrino luminosity of \(\sim10^{52}~\mathrm{erg~s^{-1}}\). 
The duration of nucleosynthesis can be arbitrarily chosen up to \(1\,\mathrm{s}\), which is longer than the typical duration of explosive nucleosynthesis, because the history of the abundance of each nucleus is maintained. 
In total, we calculate \(\sim 2000\) models with different \(T\), \(\rho\), and \fnu.

\begin{table}[ht!]
\caption{Parameter Sets for Nucleosynthesis Calculations}
\label{table:parameters}
\begin{tabular}{ccccc}\hline
 & Temperature  & Density & Neutrino Flux & Duration of Nucleosynthesis \\ 
 & \(\mathrm{[GK]}\) & \(\mathrm{[g~cm^{-3}]}\) & \(\mathrm{[erg~cm^{-2}~s^{-1}]}\) &  \(\mathrm{[s]}\) \\ \hline
Parameter Range & \(1.0\,\mathchar`-\,7.0\) & \(10^5\,\mathchar`-\,10^8\) & \(2.65\times10^{31}\,\mathchar`-\,2.65\times10^{37}\) & \(10^{-3}\,\mathchar`-\,1\)  \\ \hline
Step Size & \(\Delta T =0.1\) & \(\Delta(\log\rho) = 1\) & \(\Delta(\log F_\nu) = 1\) & 
\\ \hline
\end{tabular}
\end{table}

We treat the neutrino processes in the nucleosynthesis calculation by following the proscription of \citet{Yoshida08}. 
The neutrino spectra are assumed to have a Fermi-Dirac distribution with zero chemical potential. The average neutrino temperatures of the electron neutrino \(\nu_e\), electron anti-neutrino \(\bar{\nu_e}\), and other neutrinos and anti-neutrinos \(\nu_x\) are set as \(4\,\mathrm{MeV}/k_\mathrm{B}\), \(4\,\mathrm{MeV}/k_\mathrm{B}\), and \(6\,\mathrm{MeV}/k_\mathrm{B}\), respectively. The energy flux is equally distributed among the flavors. 
Since we assume constant neutrino fluxes, the total neutrino exposure \(\sigma_\nu\) is defined as 
\(\sigma_\nu = F_\nu\cdot t_\mathrm{nuc}\).

\subsection{Nucleosynthesis Calculation}

Nucleosynthesis is calculated using a nuclear network comprising 809 nuclear species, as summarized in Table~\ref{tab:nuclei}. The thermonuclear reaction rates are taken from the database of JINA Reaclib database \citep{Cyburt10}, except for the weak interactions, \(\beta\)-decay and electron capture reactions, which are obtained from \citet{ffn82, ffn85, Oda94, Langanke01}. 
We note that neutral current reactions such as that described by equation (1) in \citet{Woosley90} are the most important for the synthesis of Sc and V in the neutrino process because they are mainly synthesized by both of proton and neutron capture reactions. The neutral current reactions contribute little to the heating of materials because those are scattering processes.


\begin{table}[ht!]
\caption{Nuclear species in the reaction network}
\begin{tabular}{cccccccccc}\hline\hline
Isotope & \(A\) & Isotope & \(A\) & Isotope & \(A\) & Isotope & \(A\) & Isotope & \(A\) \\ \hline
n  & \(1\)         & Ne & \(18\)-\(26\) & Ca & \(38\)-\(52\) & Zn & \(59\)-\(78\)  & Zr & \(74\)-\(105\) \\
H  & \(1\)-\(3\)   & Na & \(21\)-\(28\) & Sc & \(40\)-\(55\) & Ga & \(60\)-\(81\)  & Nb & \(80\)-\(107\) \\
He & \(3\)-\(4\)   & Mg & \(22\)-\(31\) & Ti & \(42\)-\(57\) & Ge & \(59\)-\(84\)  & Mo & \(79\)-\(110\) \\
Li & \(6\)-\(7\)   & Al & \(25\)-\(34\) &  V & \(44\)-\(60\) & As & \(64\)-\(86\)  & Tc & \(85\)-\(113\) \\
Be & \(7\)-\(9\)   & Si & \(26\)-\(36\) & Cr & \(46\)-\(63\) & Se & \(65\)-\(89\)  & Ru & \(84\)-\(115\) \\
B  & \(8\)-\(13\)  &  P & \(27\)-\(39\) & Mn & \(48\)-\(65\) & Br & \(68\)-\(92\)  & Rh & \(89\)-\(118\) \\
C  & \(11\)-\(15\) &  S & \(30\)-\(42\) & Fe & \(50\)-\(68\) & Kr & \(66\)-\(94\)  & Pd & \(89\)-\(121\) \\
N  & \(13\)-\(18\) & Cl & \(32\)-\(44\) & Co & \(51\)-\(71\) & Rb & \(72\)-\(96\)  &    &           \\
O  & \(14\)-\(21\) & Ar & \(34\)-\(47\) & Ni & \(54\)-\(73\) & Sr & \(69\)-\(100\) &    &           \\
F  & \(17\)-\(23\) &  K & \(36\)-\(50\) & Cu & \(56\)-\(76\) &  Y & \(76\)-\(102\) &    &         \\ \hline  
\end{tabular}
\label{tab:nuclei}
\end{table}





\subsection{Observational Data}

We use the data for VMP stars summarized in the Stellar Abundances for Galactic Archaeology (SAGA) database \citep{suda08, suda11, Yamada13, suda17}. 
Because we focus on the abundance ratios among Sc, Ti, and V, as well as the incomplete Si burning, we adopt the abundance ratios [(Sc, V)/Ti] and compare the observations with the results of nucleosynthesis models (Section~\ref{sec:mod}). 
The gray scale in Figure \ref{fig:saga_wnu-wonu} shows the distribution of the [Sc/Ti] and [V/Ti] of the VMP stars. 
In this paper, the VMP stars with suspiciously high abundance ratios ([(Sc, V)/Ti]\,\(\geq +2.0\))\footnote{We also exclude VMP stars reported in the same literature \citep{Kipper94, Placco14, Sakari18}.} or with only upper limits are excluded and 441 VMP stars are used for comparison. 
Gaussian distributions on [Sc/Ti]-[V/Ti] plane with spreads corresponding to the error bars are adopted for the stars\footnote{We assume the errors of \(0.2\)~dex for stars without errors.} and summed up for all of the stars. 
We adopt a 90\% contour in the [Sc/Ti]-[V/Ti] plane to diagnose the consistency between the observations and the models.




\section{Results} \label{sec:results}




\begin{figure}[ht!]
    \centering
    \includegraphics[width=0.6\linewidth]{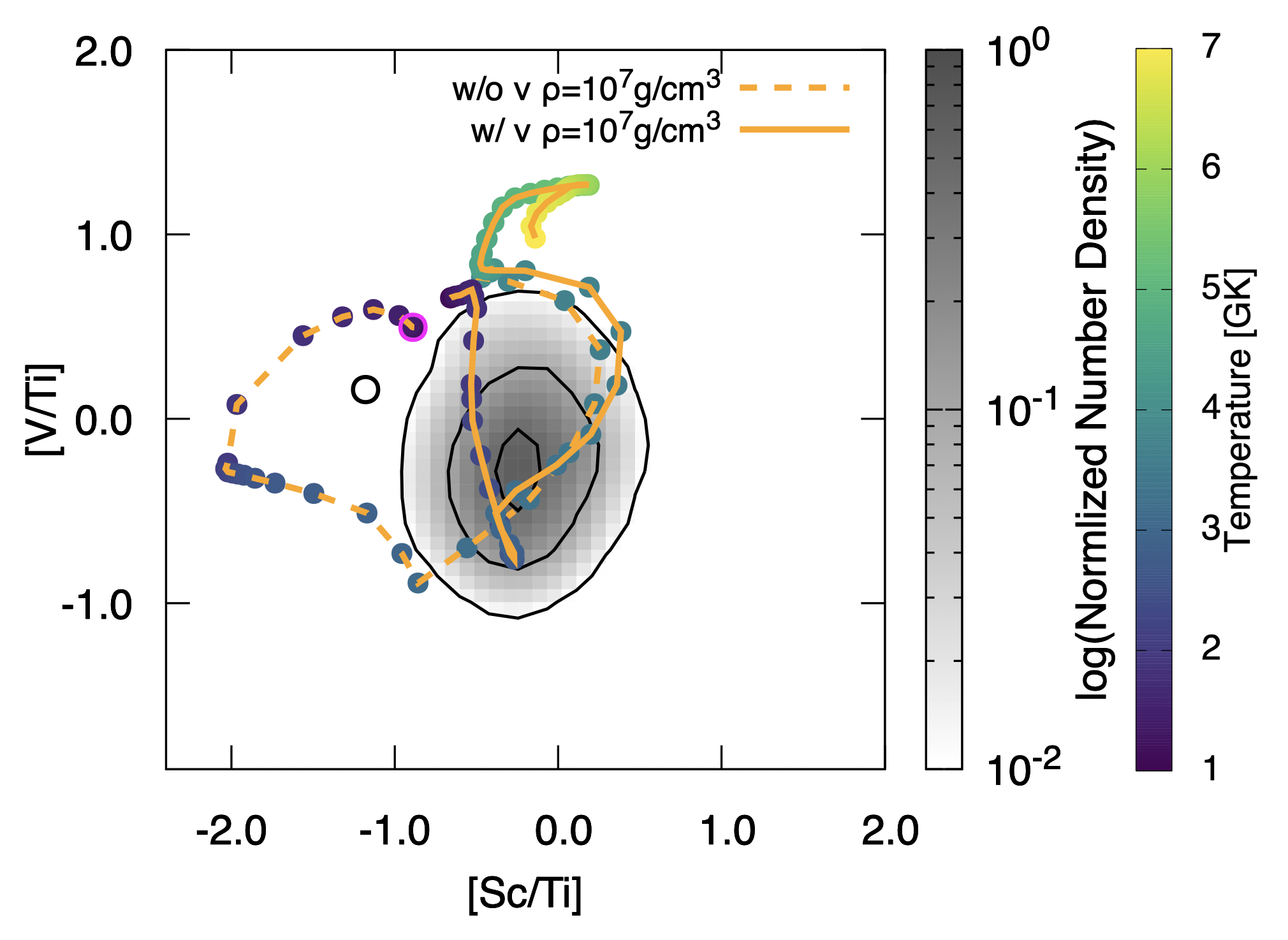}
    \caption{Gray scale is the normalized number density of the VMP stars as functions of [Sc/Ti] and [V/Ti]. The abundance ratios of 441 VMP stars are taken from the SAGA database \citep{suda08, suda11, Yamada13, suda17}. The contours represent 50\%, 90\%, and 99\% of the total from the inside to outside. The [Sc/Ti] and [V/Ti] of the models with constant \(\rho\)\,(\(=10^{7}\,\mathrm{g~cm^{-3}}\)), \(F_\nu\)\,(\(=2.65\times10^{36}\,\mathrm{erg~cm^{-2}~s^{-1}}\)), \(t_\mathrm{nuc}\)\,(\(=0.1\,\mathrm{s}\)), and various \(T\) (filled circles connected with lines). The abundance ratios in an adopted progenitor star and an explosion model of the same progenitor star without neutrino exposure calculated in \citet{Tominaga07} are shown using a magenta open circle and a black circle, respectively. The models with the same \(\rho\), \fnu, and \tnuc\ are connected lines. The colors of the circles and lines represent \(T \) and \(\rho\). }
    \label{fig:saga_wnu-wonu}
\end{figure}

Figure \ref{fig:saga_wnu-wonu} illustrates the results of nucleosynthesis calculations with and without neutrinos in comparison with the observed abundance ratios of the VMP stars.
The models in the figure have \((F_\nu,~t_\mathrm{nuc}) = (2.65\times10^{36}\,\mathrm{erg~cm^{-2}~s^{-1}},~0.1\,\mathrm{s})\). Each point shows the resultant [Sc/Ti] and [V/Ti] ratios of models with different \(T\) and constant density (\(\rho=10^7\,\mathrm{g~cm^{-3}}\)).
As shown in Figure \ref{fig:saga_wnu-wonu}, most of the models without neutrinos are not located in the 90\% contour because of the underproduction of Sc in \(T<3.2\,\mathrm{GK}\) and the overproduction of V in \(T<2.0\,\mathrm{GK}\) and \(T>3.5\,\mathrm{GK}\). In contrast, the Sc is enhanced in the range of \(T<3.2\,\mathrm{GK}\) in the models with neutrinos, resulting in the successful reproduction of the observational [(Sc, V)/Ti] ratios of the models with \(T=2.0\,\mathchar`-\,3.4\,\mathrm{GK}\). 
At high temperatures (\(T>4.0\,\mathrm{GK}\)), the abundance ratios are almost the same in both models, with and without neutrinos, because a nuclear statistical equilibrium (NSE) is established.

\begin{figure}[ht!]
    \centering
    \includegraphics[width=1\linewidth]{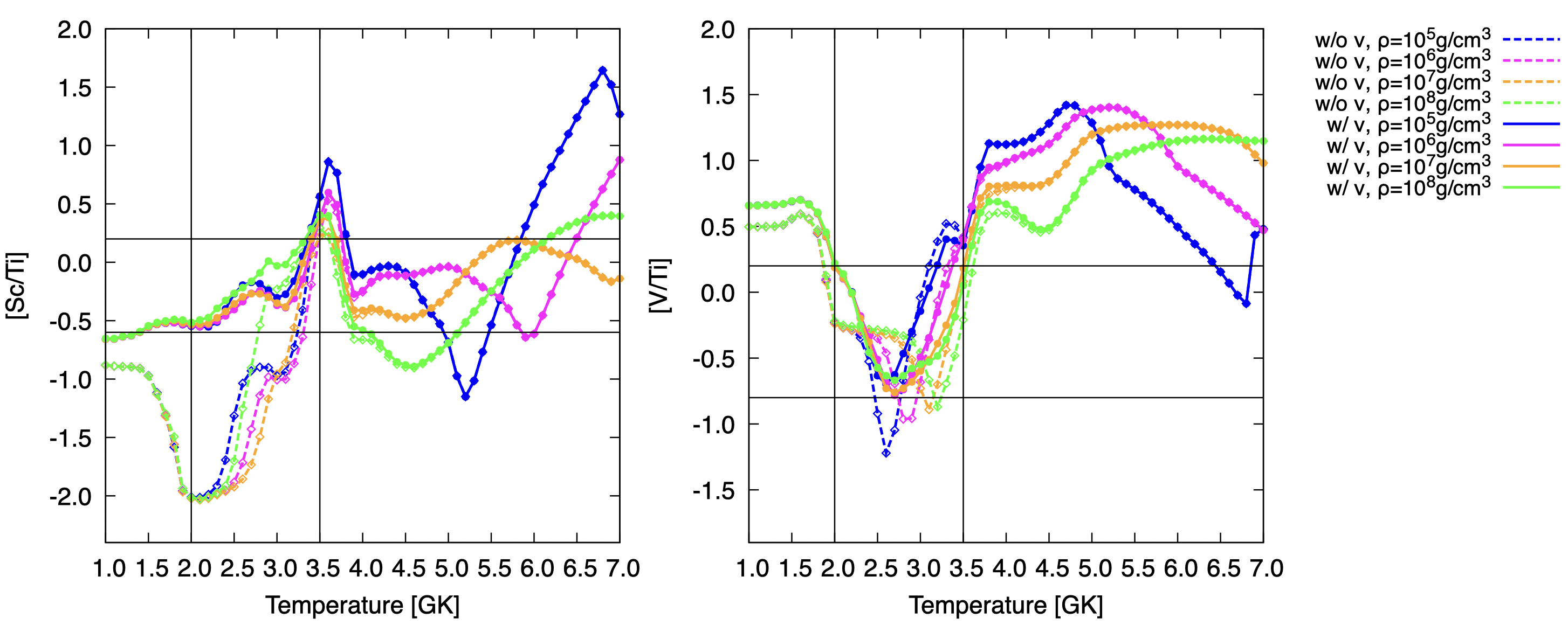}
    \caption{[Sc/Ti] and [V/Ti] of the models with \(F_\nu=2.65\times10^{36}\,\mathrm{erg~cm^{-2}~s^{-1}}\), \(t_\mathrm{nuc}=0.1\,\mathrm{s}\) and various \(T\) and \(\rho\). Colors of lines correspond to density, from \(\rho=10^5\,\mathrm{g~cm^{-3}}\) to \(10^8\,\mathrm{g~cm^{-3}}\). Solid and dashed lines indicate the models with and without neutrinos, respectively. The black lines show the  [Sc/Ti] and [V/Ti] values at the boundary of 90\% contour line.
    }
    \label{fig:neu36t01s_T-Sc_V}
\end{figure}

\begin{figure}[ht!]
    \centering
    \includegraphics[width=1\linewidth]{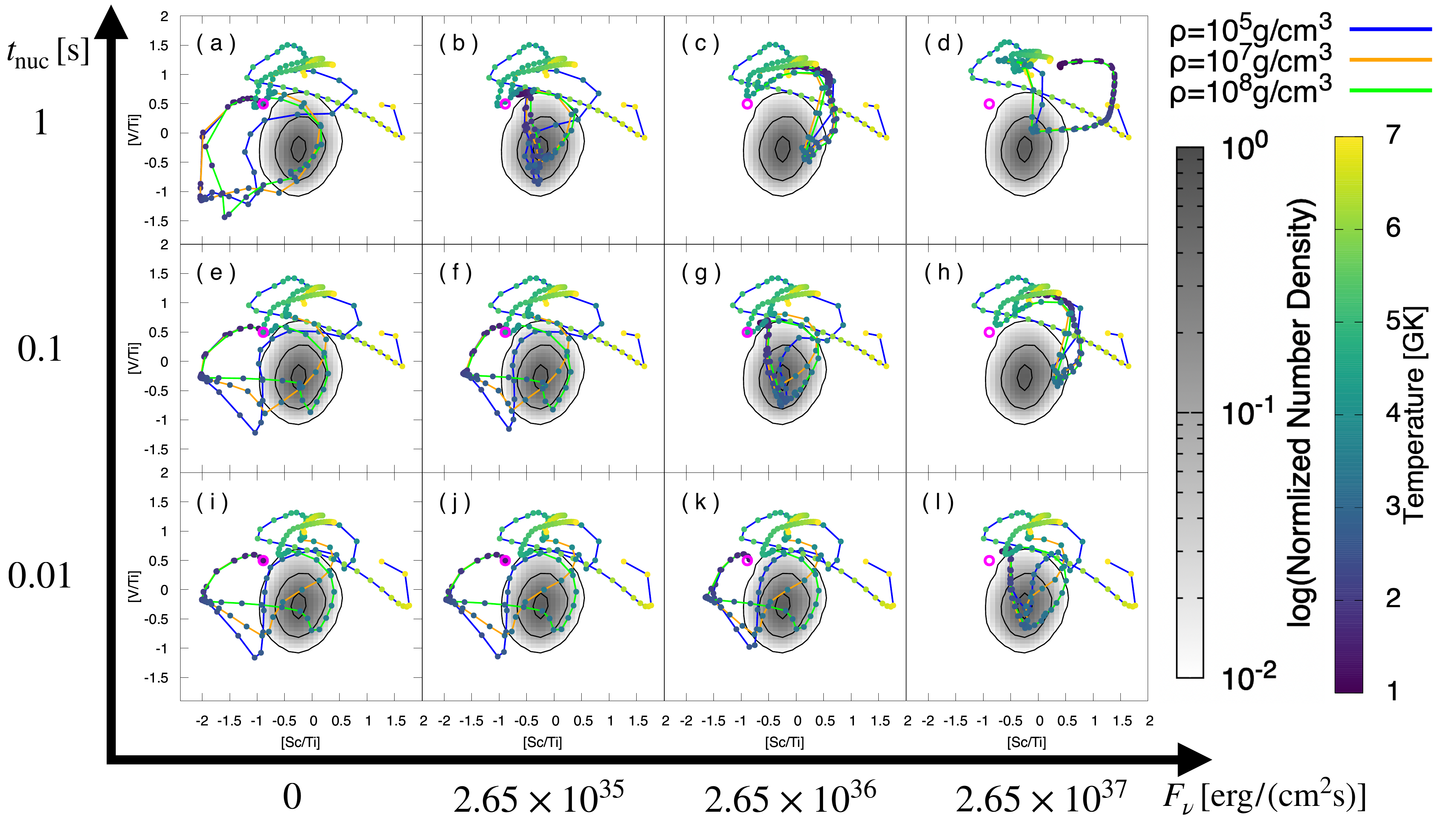}
    \caption{Abundance ratios [Sc/Ti] and [V/Ti] of the models with various \(T\), \(\rho\), \fnu, and \tnuc\ (filled circles connected with lines). The same \fnu\ and \tnuc\ are adopted for the models in each panel. The colors of the circles and lines represent \(T\) and \(\rho\), respectively. The magenta open circles, background gray scale, and contours are the same as those in Figure \ref{fig:saga_wnu-wonu}.}
    \label{fig:nonneu_neu_12}
\end{figure}

Figure \ref{fig:neu36t01s_T-Sc_V} shows the temperature dependence of [Sc/Ti] and [V/Ti] in the models with \(F_\nu=2.65\times10^{36}\,\mathrm{erg~cm^{-2}~s^{-1}}\) and \(t_\mathrm{nuc}=0.1\,\mathrm{s}\). The left panel illustrates that the neutrino processes enhance the [Sc/Ti] ratios at \(T \leq3.4\,\mathrm{GK}\). 
However, the influence of neutrinos on V synthesis is weaker than that on Sc synthesis. Independent of the neutrino processes, the resultant [V/Ti] ratios in most of models with \(2.0\,\mathrm{GK} \leq T \leq 3.2\, \mathchar`-\,3.5\,\mathrm{GK}\) are consistent with those seen in the observations. Although the upper limit of the temperature slightly depends on the density, a range of \(2.0\,\mathrm{GK} \leq T \leq 3.2\,\mathrm{GK}\) is required to reproduce the abundances ratio of VMP stars in the models with \(F_\nu=2.65\times10^{36}\,\mathrm{erg~cm^{-2}~s^{-1}}\) and \(t_\mathrm{nuc}=0.1\,\mathrm{s}\). 

We next explain below how this condition changes for other parameters of the neutrino flux and duration of nucleosynthesis.
Figure \ref{fig:nonneu_neu_12} summarizes the abundance ratios of all the models and compares them with the observed abundance ratios of the VMP stars.
The models in each panel have the same \fnu\ and \tnuc. The value of neutrino exposure \(\sigma_\nu\) increases from the bottom left panel toward the top right panel. Each panel shows the resulting [Sc/Ti] and [V/Ti] of the models for various \(T\) and \(\rho\). 


Panels (a), (e), and (i) show the models without neutrino irradiation. The models exhibit an underproduction of Sc at low temperature (\(T\leq2.0\)~GK) or overproduction of V at high temperature (\(T\geq5.0 \)~GK). Although some models with \(T\sim 3.0\)~GK are located in the 90\% contour, the majority of the models cannot reproduce the observations because of their strong dependence on \(T\). This illustrates the difficulty of reproducing of [(Sc, V)/Ti] of the VMP stars without neutrino irradiation, because the temperature range of reproduction in the models without neutrino irradiation is very narrow comparing to the temperature range at which explosive nucleosynthesis takes place.


In the three columns on right, panels (b), (c), (d), (f), (g), (h), (j), (k), and (l), show the models with neutrino irradiation. The [Sc/Ti] is higher for the models with greater neutrino exposure. The observed abundance ratios are reproduced well by the models with low temperatures of \(T=2.0\,\mathchar`-\,3.2\)~GK, as seen in panel (b), (g), (l) with \((F_\nu,~t_\mathrm{nuc}) = (2.65\times10^{35}~\mathrm{erg~cm^{-2}~s^{-1}},~1~\mathrm{s}),\,(2.65\times10^{36}~\mathrm{erg~cm^{-2}~s^{-1}},~0.1~\mathrm{s})\), and \(~(2.65\times10^{37}~\mathrm{erg~cm^{-2}~s^{-1}},~0.01~\mathrm{s})\). This indicates that neutrino exposure \(\sigma_\nu \sim10^{35}~\mathrm{erg~cm^{-2}}\) is required to reproduce the [(Sc, V)/Ti] of the VMP stars, independent of the values of \fnu\ and \tnuc.  
If the neutrino exposure is smaller or larger than \(\sigma_\nu \sim10^{35}~\mathrm{erg~cm^{-2}}\), the resultant [Sc/Ti] is lower or higher than the observations, respectively.
Furthermore, the width of the distribution of the observed abundance ratios is consistent with the scatter of the models with \(\sigma_\nu \sim10^{35}~\mathrm{erg~cm^{-2}}\) and \(T=2.0\,\mathchar`-\,3.2\)~GK. 

In the models with high temperatures at \(T>5.0\)~GK, NSE is established. The models do not reproduce the observed abundance ratios, owing to the overproduction of V with high \(\rho~(\geq 10^6~\mathrm{g~cm^{-3}})\) and the overproduction of Sc with low \(\rho~(=10^5~\mathrm{g~cm^{-3}})\). The high [Sc/Ti] ratio in the models with high \(T\) and low \(\rho\) results from the high entropy. 
If NSE is established, the abundances are determined by \(T\), \(\rho\), and \(Y_e\). Here, because the electron fraction evolves over time owing to neutrino irradiation, the resultant abundance ratios depend on neutrino exposure and slightly change even in NSE. 


In the models with low temperatures (\(T\simeq1.0\)~GK), explosive nucleosynthesis proceeded slowly. As a result, the initial abundance ratios are maintained in the models with small neutrino exposure (\(\sigma_\nu \lesssim 10^{34}~\mathrm{erg~cm^{-2}}\)). If the neutrino exposure is large (\(\sigma_\nu\gtrsim10^{35}~\mathrm{erg~cm^{-2}} \)), the neutrino process slightly enhances the [Sc/Ti] because of the neutral current reaction; however, the abundance ratio is inconsistent with the observed abundance ratios because of the high [V/Ti]~\(\simeq+1.0\).

%


Consequently, we conclude that the models with \(\sigma_\nu \sim10^{35}\,\mathrm{erg~cm^{-2}}\) and \(T=2.0\,\mathchar`- \,3.2\)~GK reproduce the observed abundance ratios [Sc/Ti] and [V/Ti].

\section{Discussion} \label{sec:discussion}

In this section, we discuss the physical mechanisms that realize the above conditions to reproduce the abundance ratios among Sc, Ti, and V, i.e., \(\sigma_\nu \sim10^{35}~\mathrm{erg~cm^{-2}}\) and \(T=2.0 \, \mathchar`-\,3.2\)~GK. 
Assuming the typical neutrino luminosity \(L_\nu\sim10^{52}~\mathrm{erg~s^{-1}}\) of CCSNe, the requirement for the neutrino exposure of \(\sim10^{35}\)~erg~cm$^{-2} \) is realized if the materials drift at \(r\sim10^7~\mathrm{cm}\) for \(\sim0.01~\mathrm{s}\) or at \(r\sim10^8~\mathrm{cm}\) for \(\sim1~\mathrm{s}\). 

Firstly, we consider a one-dimensional model. In the radiation-dominated environment behind the shock wave, the following relation exists between the post-shock temperature and the radius of the shock wave: \(r \simeq 3.16\times10^9{(E_\mathrm{exp}/1~\mathrm{B})}^{1/3}{(T/10^9~\mathrm{K})}^{4/3}~\mathrm{cm}\), where \(E_\mathrm{exp}\) is the explosion energy.
Assuming a typical explosion energy of \(0.6~\mathrm{B}\), the peak temperature reaches \(3.0~\mathrm{GK}\) at \(r\simeq1.15\times10^9~\mathrm{cm}\) where the neutrino flux is too low to satisfy the neutrino exposure condition. This is why Sc is under-produced in the case of typical neutrino energies \citep{Yoshida08}.

In typical three-dimensional simulations, the calculations are performed for less than \(1\,\mathrm{s}\) after the core bounce \citep[e.g.][]{Nakamura25}, making it difficult to investigate the particle dynamics in the regions far from the center.
However, the long-term three-dimensional simulations, for example, \citet{Sieverding23a, Wang24a} showed that some mass particles experience a few seconds scale turbulent motion and exhibit non-monotonic thermal evolutions. 
During the turbulent motion, some particles experience the temperature of \(\sim2.0\,\mathrm{GK}\) and the radii of \(\sim10^8~\mathrm{cm}\) \citep[see Figure 6 and Figure 7 in][]{Sieverding23a}. These particles could be likely to satisfy the conditions on \(T \) and \snu. Their density during the turbulent motion is also around a few \(\times10^5~\mathrm{g~cm^{-3}}\) which is consistent with the values adopted in this study. 
Although further investigations are needed to constrain the total mass of such particles, their presence in the three-dimensional simulations implies that the requirements on temperature and neutrino exposure found in this study are feasible.



The progenitor star used in \citet{Sieverding23a} experienced a convective oxygen-shell burning and violent Ne-O shell merger. This results in the aspherical structure of the progenitor model \citep{Yadav20}. This implies that the initial asphericity in the progenitor is the key to the turbulent structure a few seconds after the core bounce. Hence, it is important to self-consistently calculate the nucleosynthesis, including Sc, Ti, and V in the explosions of aspherical and metal-free progenitor models.






\section{Conclusion} \label{sec:conclusion}

We investigate the synthesis of Sc, Ti, and V in CCSNe and identify the requirement on neutrino exposure and temperature to reproduce the abundance ratios among Sc, Ti, and V in the VMP stars. We adopt the \(13M_\odot\) zero-metal progenitor model \citep{Umeda00} and calculate nucleosynthesis including neutrino processes without any explosion models in the layer of \(1.5\,\mathchar`-\,1.7 M_\odot\). This layer is adopted because Sc, Ti, and V are mainly synthesized via the explosive Si burning in this layer \citep{Tominaga07}. 
The nucleosynthesis results are compared with the abundance ratios in the VMP stars which are obtained from the SAGA database, and we find the requirements on neutrino exposure of \(\sim 10^{35}\)~erg~cm\(^{-2}\) and temperature of \(2.0\,\mathchar`-\,3.2\)~GK, which are determined by [Sc/Ti] and [V/Ti], respectively.

Additionally, we consider whether the requirements are realized in previous explosion simulations by calculating nucleosynthesis without assuming any explosion models. In one-dimensional models, materials experience \(T\sim 3.0\)~GK at \(r\sim10^9~\mathrm{cm}\) where the neutrino flux is too low to reach the condition of neutrino exposure. In long-term three-dimensional simulations, some particles undergo non-monotonic thermal history due to turbulent motion \citep{Sieverding23a, Wang24a}. \citet{Sieverding23a} adopted an aspherical progenitor model, having experienced a convective oxygen-shell burning and violent Ne-O shell merger \citep{Yadav20}. This implies that the turbulent motion during the explosion caused by the initial asphericity could be crucial for satisfying the requirements on neutrino exposure and temperature for the reproduction of the abundance ratios among Sc, Ti, and V. More detailed investigations on the total mass of particles which satisfy the conditions are desired.



In this study, for the first time, we identify the local physical conditions on the neutrino exposure and temperature to reproduce the observed abundance ratios among Sc, Ti, and V. 
Here, we focus on the abundances ratios [(Sc, V)/Ti] and calculate nucleosynthesis without assuming any hydrodynamical model to solve the Sc, Ti, and V problem. Further investigations are necessary as follows; 
(1) Self-consistent nucleosynthesis calculations based on long-term explosion simulations of multi-dimensional metal-free progenitor models should be performed; 
(2) We did not discuss the abundance ratios among Sc, Ti, and V to Fe because these ratios depend on additional factors, such as the geometry of the explosion. The above self-consistent simulations are needed to assess the abundance ratios between the elements synthesized in the different layers.


\begin{acknowledgments}
Numerical computations were carried out on PC cluster at Center for Computational Astrophysics, National Astronomical Observatory of Japan. This study was supported by JSPS KAKENHI (21H04499, 23H04894, 25H02196, 25K01046). This work was partially supported by Overseas Travel Fund for Students (2023, 2024, and 2025) of Astronomical Science Program, The Graduate University for Advanced Studies, SOKENDAI.

\end{acknowledgments}

\bibliography{sample7}{}
\bibliographystyle{aasjournal}



\end{document}